\documentclass[11pt]{article}
\usepackage[utf8]{inputenc}

\usepackage{varioref}
\usepackage{amsmath,amsfonts,amsthm,mathrsfs}
\usepackage{mathabx}
\usepackage[normalem]{ulem}
\usepackage{multirow,color,graphics}
\usepackage{amsmath}
\usepackage{bbold}
\usepackage{amsfonts}
\usepackage{amssymb}
\usepackage{jheppub}
\usepackage{enumerate}
\usepackage{fancyvrb}
\usepackage{verbatim}
\usepackage{wrapfig}
\usepackage{appendix}
\usepackage{amstext}
\usepackage{amssymb}
\usepackage{graphicx}
\usepackage{color}
\usepackage{varioref}
\usepackage{multirow,graphics}
\usepackage{epstopdf}
\usepackage{mathtools}
\usepackage{array}
\newcommand{\nn}{\nonumber}

\numberwithin{equation}{section}

\def\[{\left[}
\def\]{\right]}
\def\({\left(}
\def\){\right)}
\def\<{\left<}
\def\>{\right>}

    \newcommand{\beq}{\begin{equation}}
    \newcommand{\eeq}{\end{equation}}
    \newcommand\beqa{\begin{eqnarray}}
    \newcommand\eeqa{\end{eqnarray}}
\newcommand\bea{\begin{array}}
\newcommand\eea{\end{array}}

\newcommand{\bQ}{{\bf Q}}

\newcommand{\bP}{{\bf P}}
\newcommand{\bp}{{\bf p}}

\newcommand{\la}[1]{\label{#1}}
\newcommand{\eq}[1]{(\ref{#1})}

    \def\bQ{{\bf Q}}
    \def\bP{{\bf P}}

\definecolor{cadmiumgreen}{rgb}{0.0, 0.42, 0.24}

\makeatletter
     \@ifundefined{usebibtex}{} {}
\makeatother

    \def\bQ{{\bf Q}}

        \def\bP{{\bf P}}

     \def\Tr{\text{Tr}}

\newcommand{\cN}{\mathcal{N}}

\newcommand{\cO}{\mathcal{O}} 

\renewcommand{\<}{\langle} 
\renewcommand{\>}{\rangle}

\title{
Exploring the ground state spectrum of $\gamma$-deformed N=4 SYM}

\author[a,1]{~~Fedor Levkovich-Maslyuk\note{Also at Institute for Information Transmission Problems, Moscow 127994, Russia},}
\author[b]{~~Michelangelo Preti}

\affiliation[a]{Laboratoire de Physique de l’Ecole Normale Superieure, ENS, Universite PSL, CNRS, Sorbonne Universite, Universite de Paris, F-75005 Paris, France }

\affiliation[b]{Nordita, KTH Royal Institute of Technology and Stockholm University, Roslagstullsbacken 23, SE-10691 Stockholm, Sweden}

\emailAdd{fedor.levkovich$\bullet$gmail.com} 
\emailAdd{michelangelo.preti$\bullet$gmail.com} 

\abstract{We study the $\gamma$-deformation of the planar $\cN=4$ super Yang-Mills theory which breaks all supersymmetries but is expected to preserve integrability of the model. We focus on the operator $\Tr(\phi_1\phi_1)$ built from two scalars, whose integrability description has been questioned before  due to contributions from double-trace counterterms. We show that despite these subtle effects, the integrability-based Quantum Spectral Curve (QSC) framework works perfectly for this state and in particular reproduces the known 1-loop prediction. This resolves an earlier controversy concerning this operator and provides further evidence that the $\gamma$-deformed model is an integrable CFT at least in the planar limit. We use the QSC to compute the first 5 weak coupling orders of the anomalous dimension analytically, matching known results in the fishnet limit, and also compute it numerically all the way from weak to strong coupling. We also utilize this data to extract a new coefficient of the beta function of the double-trace operator couplings.}

\begin{document}

\maketitle

\section{Introduction}

In recent years, powerful methods based on integrability have led to the calculation of a wide range of observables in planar $\cN=4$ SYM theory at the non-perturbative level \cite{Beisert:2010jr}. The success of the integrability program has motivated a search for other solvable models in 4d with less symmetry than the original theory, with the goal of getting closer to realistic models as well as better understanding the mechanisms behind integrability in general.

Remarkably, there exists a deformed version of SYM, known as the $\gamma$-deformation \cite{Lunin:2005jy,Frolov:2005dj,Beisert:2005if,Frolov:2005iq}, which no longer has any supersymmetry but appears to retain integrability as well as conformal invariance. It is a 3-parametric family of theories obtained by inserting extra constant phase factors into the Lagrangian depending on three angles $\gamma_1,\gamma_2,\gamma_3$. The particular case of $\gamma_1=\gamma_2=\gamma_3$ corresponds to the $\beta$-deformation and has also been much studied. In the dual string model these angles parametrize a TsT transformation of the background, and in the integrability description they correspond to twisted boundary conditions. Many of the powerful integrability techniques developed for the spectrum of anomalous dimensions in the original model, such as the asymptotic Bethe ansatz \cite{Beisert:2005fw}, Y-system \cite{Gromov:2009tv}, Thermodynamic Bethe Ansatz (TBA) \cite{Gromov:2009bc,Bombardelli:2009ns,Arutyunov:2009ur} and finally the Quantum Spectral Curve (QSC) \cite{Gromov:2013pga}, have been translated to the deformed version in respectively \cite{Beisert:2005if}, \cite{Gromov:2010dy}, \cite{deLeeuw:2012hp} and \cite{Kazakov:2015efa} (see also the review \cite{Zoubos:2010kh}). 

Despite this progress, it was realized that a special subset of the usual single-trace operators  exhibits rather peculiar features \cite{Fokken:2013aea,Fokken:2013mza,Fokken:2014soa} (see also e.g. \cite{Tseytlin:1999ii,Dymarsky:2005nc,Pomoni:2008de}). Namely, even in the strict planar limit their anomalous dimensions receive contributions from double-trace counterterms that are needed to render the 2-point function finite. In turn this puts into question the conformal invariance of the theory, since the counterterms have nontrivial beta-functions. While these subtle effects have an impact only on a restricted set of states, it has remained an important problem to clarify their properties and in particular to understand whether integrability is preserved for these special states.

In this paper we focus on the simplest of such states, namely the operator built from two scalars $\Tr(\phi_1\phi_1)$. In the original theory all operators $\Tr(\phi_1^J)$ are protected BMN vacua with dimensions $\Delta=J$, but in the $\gamma-$deformed model they acquire nontrivial anomalous dimensions. The states with $J\geq 3$ are well described by usual integrability methods \cite{Ahn:2011xq,deLeeuw:2012hp} and do not feel the double-trace effects discussed above. Yet, suprisingly, for $J=2$ the integrability-based Thermodynamic Bethe Ansatz appears to give a divergent result at weak coupling, as found in \cite{Ahn:2011xq}  (see also \cite{deLeeuw:2011rw}). At the same time, a careful diagrammatic computation at weak coupling  \cite{Fokken:2014soa} revealed a finite but unexpected contribution of order $g^2$ to its anomalous dimension termed `prewrapping', originating in the coupling to double-trace operators and apparently not captured by integrability. Later it was also suggested in \cite{Kazakov:2015efa} that the more advanced Quantum Spectral Curve may exhibit a singular behavior for this state.

The consequences of the double-trace running couplings were later understood much better in \cite{Grabner:2017pgm,Gromov:2018hut} on the example of the fishnet theory \cite{Gurdogan:2015csr,Caetano:2016ydc}, which is a further deformation obtained by sending the coupling $g$ to zero while $\gamma_j$ are formally sent to $+i\infty$ such that the combinations $\xi_j=ge^{-i\gamma_j/2}$ are held fixed and play the role of effective couplings. It was shown (at least perturbatively) that in this model the RG flow brings the double trace couplings to a fixed point where they are determined in terms of the original 't Hooft coupling and the theory becomes a true CFT, albeit a non-unitary one. Most importantly, all evidence shows that precisely at these fixed points the theory is integrable and the QSC captures its spectrum \cite{Grabner:2017pgm,Gromov:2017cja}. In particular, the $J=2$ scaling dimension has been computed analytically to all loops both directly \cite{Grabner:2017pgm} and from the QSC \cite{GrabnerToapp}
(for the case when $\xi_1=\xi_2=0$ while $\xi_3$ is arbitrary). As discused in \cite{Grabner:2017pgm,Gromov:2018hut}, it is reasonable to expect that the full $\gamma$-deformed model should display a similar behavior, becoming a conformal and integrable gauge theory at the fixed points while potentially losing unitarity.

%Examples of integrable QFTs in more than two dimensions with a known Lagrangian are very rare. The known cases, as $\mathcal{N}=4$ SYM in 4d and ABJM in 3d possess gauge symmetry and a large amount of supersymmetry that, together with the AdS/CFT string description, they have been long believed to be a prerequisite for their quantum integrability. The $\gamma$-deformed theory we study, despite the absence of any supersymmetry and the non-trivial RG-flow it remains integrable. The bill to pay for this surprising feature is the loss of unitarity. As a consequence, the conformal fixed points become imaginary and the anomalous dimension of twist-two operators $\gamma=\Delta-2$ is purely imaginay.

Despite this progress, in the parent $\gamma$-deformed theory the $J=2$ anomalous dimension has never been computed from integrability, and whether this can be done at all has remained an open question. In fact, as recently as in \cite{Marboe:2019wyc} it was suggested that the QSC for this state, while presumably giving a finite result, may display some extremely unusual features, such as intermediate quantities having an expansion in not only even but also odd powers of the coupling.

Here we present results that should settle the controversy around the integrability description of this $J=2$ operator. We compute its anomalous dimension in a variety of regimes by solving the Quantum Spectral Curve equations,  and demonstrate that they give a perfectly finite result and do not reveal any unusual properties. The main complication are technical difficulties which we overcome by carefully applying all the experience developed by now with the undeformed QSC. In particular, to get the result even at 1 loop we need to solve the QSC perturbatively to a rather high order due to some cancellations. However, when the dust settles we precisely reproduce the diagrammatic 1-loop result of \cite{Fokken:2014soa} at any value of the deformation parameteres. This demonstrates that, like in the fishnet theory, the QSC appears to incorporate automatically all the double trace contributions to the anomalous dimension! In addition, we computed the scaling dimension from the QSC to 5 loops, the result being
%\beqa
%    \Delta&=&2+8 i S_- S_+{\color{blue}g^2}
%    +
%     0\times {\color{blue}g^4}
%    \\ \nn &+&
%     \[-96   \left(S_-^2+S_+^2\right)  \zeta_3-64  S_-^2 S_+^2\]iS_-S_+{\color{blue}g^6}
%     \\ \nn
%    &+&\[
%     1024  S_-^2 S_+^2 \zeta_3+1280   \left(S_-^2+S_+^2\right)  \zeta_5 \]iS_-S_+{\color{blue}g^8}
%    \\ \nn
%    &+&\[1280  S_-^2 S_+^2\left(3 S_-^2+3 S_+^2-13\right) \zeta_5+2304  S_-^2 S_+^2 \left(S_-^2+S_+^2\right)
%   \zeta_3
%      \right. \\ \nn && \left.
%   -15680   \left(S_-^2+S_+^2\right)
%   \zeta_7
%   +576  \left(3 S_-^4-14
%   S_+^2 S_-^2+3 S_+^4\right) (\zeta_3)^2+
%   1792  S_-^4 S_+^4\]iS_-S_+{\color{blue}g^{10}}
%   \\ \nn 
%   &+&{\cal O}({\color{blue}g^{12}})
%\eeqa
\beqa
\label{Dintro}
    \Delta_\pm=2 &\pm& 8 i S_- S_+{\color{blue}g^2}
    \\ \nn &\pm&
     0\times {\color{blue}g^4}
     \\ \nn &\pm&
     32iS_-S_+\[-3   \left(S_-^2+S_+^2\right)  \zeta_3-2  S_-^2 S_+^2\]{\color{blue}g^6}
     \\ \nn
    &\pm&256iS_-S_+\[
     4  S_-^2 S_+^2 \zeta_3+5   \left(S_-^2+S_+^2\right)  \zeta_5 \]{\color{blue}g^8}
    \\ \nn
    &\pm&64iS_-S_+\[28  S_-^4 S_+^4+36  S_-^2 S_+^2 \left(S_-^2+S_+^2\right)
   \zeta_3+20  S_-^2 S_+^2\left(3 S_-^2+3 S_+^2-13\right) \zeta_5
      \right. \\ \nn && \left.
   +9  \left(3 S_-^4-14
   S_+^2 S_-^2+3 S_+^4\right) (\zeta_3)^2 -245   \left(S_-^2+S_+^2\right)
   \zeta_7\]{\color{blue}g^{10}}+{\cal O}({\color{blue}g^{12}}) \ ,
\eeqa
where 
\beq
    S_\pm=\sin\(\mp\frac{\gamma_2\pm\gamma_3}{2}\) \ .
\eeq
The two possible $\pm$ signs in \eq{Dintro} correspond to choosing one of the two fixed points as expected, see section \ref{sec:gamma} for more details. This 5-loop result also agrees with all-loop predictions from the fishnet theory \cite{Grabner:2017pgm,Kazakov:2018gcy}. 
Using our data we also managed to compute the subleading order of the conformal fixed point and to reduce the computation of the $\beta$-function to a small set of Feynman diagrams. 
We also solved the QSC numerically for a wide range of the coupling.

Since the QSC itself at least in some cases can be derived from the TBA \cite{Gromov:2014caa}, these two approaches are expected to provide the same result for the spectrum. Thus it may be possible to extract the finite answer from TBA as well, by introducing a careful regularization in the computation of \cite{Ahn:2011xq}.

Let us note that examples of conformal, non-supersymmetric gauge theories in 4d with a known  Lagrangian are very rare. The $\gamma$-deformed model indeed seems to be a theory in this class, which moreover also appears to be integrable (in the planar limit) as we further confirm in this paper. The price to pay for this remarkable combination of features is the loss of unitarity. While only a restricted class of operators are sensitive to it, at the conformal fixed points some couplings become complex and the anomalous dimension of twist-two operators $\gamma=\Delta-2$ is purely imaginary as we will see explicitly.

The operator we consider is potentially one of the simplest states in the theory, as the shortest nontrivial operator whose scaling dimension is moreover known analytically in the fishnet limit \cite{Grabner:2017pgm,Kazakov:2018gcy}. We hope that the high-order results we present here may reveal extra insights into the structure of its spectrum and perhaps lead to further simplifications in its QSC description.

This paper is structured as follows. In section \ref{sec:gamma} we review in more detail the $\gamma$-deformed model and its renormalization. In section \ref{sec:qsc} we discuss its description in terms of the Quantum Spectral Curve. Then in section \ref{sec:sols} we discuss the weak coupling solution of the QSC and present our results. In the next section \ref{sec:num} we give our numerical results at finite coupling. We conclude in section \ref{sec:concl}, while the appendix contains  technical details. The paper is also accompanied by a Mathematica notebook with some QSC relations that are too lengthy for the main text.

\section{The $\gamma$-deformed $\mathcal{N}=4$ SYM theory}
\label{sec:gamma}

In this paper we study a deformation of $\mathcal{N}=4$ 
SYM obtained by replacing the ordinary product with the following associative, non-commutative $\star$-product \cite{Lunin:2005jy}
\beq\label{starprod}
A \star B\coloneqq e^{i\tfrac{\gamma_i}{2}[J_i(A)J_k(B)-J_k(A)J_i(B)]}\,A\,B \ ,
\eeq
where $J_i(\Phi)$ is the $U(1)_i$ charge of the field $\Phi$. The $\mathcal{N}=4$ SYM theory has $PSU(2,2|4)$ global symmetry. Choosing \eqref{starprod} to act upon the Cartan subalgebra of $SU(4)_R$, the R-symmetry is completely broken with a 3-parameter deformation labelled by $i,k=1,2,3$ such that $PSU(2,2|4)\rightarrow SU(2,2)\otimes U(1)^{\otimes 3}$.  

The resulting $\gamma$-deformed $\mathcal{N}=4$ Lagrangian is the following (see e.g. \cite{Fokken:2013aea})
\begin{equation}
\label{N=4SYMlagrangian}
  {\cal L}=N_c\Tr\biggl[
  -\frac{1}{4} F_{\mu\nu}F^{\mu\nu}
  -\frac{1}{2}D^\mu\phi^\dagger_iD_\mu\phi^i
  +i\bar \psi_{\dot\alpha \;  A}D^{\dot\alpha\alpha}\psi^A_{\alpha }\biggr]
+{\cal L}_{\rm int} \ ,
\end{equation}
where $i=1,2,3$\,\, $A =1,2,3,4$,
$D^{\dot \alpha \alpha}= D_\mu
(\bar \sigma^{\mu})^{\dot \alpha \alpha}$ and 
 \begin{equation}\label{Lint}
   \begin{aligned}[y]
     &\mathcal{L}_{\rm int} =N_cg_{\rm YM}\,\,\Tr\bigl[\frac{g_{YM}}{4} \{\phi^\dagger_i,\phi^i\}
     \{\phi^\dagger_j,\phi^j\}-g_{\rm YM}\,e^{-i\epsilon^{ijk}\gamma_k}
     \phi^\dagger_i\phi^\dagger_j\phi^i\phi^j\\
     &-e^{-\frac{i}{2}\gamma^-_{j}}\bar\psi^{}_{ j}\phi^j\bar\psi_{ 4}
     +e^{+\frac{i}{2}\gamma^-_{j}}\bar\psi^{}_{ 4}\phi^j\bar\psi_{ j}
     + i\epsilon_{ijk} e^{\frac{i}{2} \epsilon_{jkm} \gamma^+_m} \psi^k \phi^i \psi^{ j}\\
     &-e^{+\frac{i}{2}\gamma^-_{j}}\psi^{}_{ 4}\phi^\dagger_j\psi_{
       j}
     +e^{-\frac{i}{2}\gamma^-_{j}}\psi^{}_{j}\phi^\dagger_j\psi_{
       4}
     + i\epsilon^{ijk} e^{\frac{i}{2} \epsilon_{jkm} \gamma^+_m} \bar\psi_{ k} \phi^\dagger_i \bar\psi_j\bigr]\hfill\, ,
   \end{aligned}
 \end{equation}
where the summation is assumed w.r.t. doubly and triply repeating indices. In \eqref{Lint} we suppress the  Lorentz indices assuming the contractions between fermions $(\psi_i)^\alpha (\psi_j)_\alpha$ and  $(\bar\psi_i)_{\dot\alpha} (\bar\psi_j)^{\dot\alpha}$. We also introduce the following notation for the twists
\beq\label{gammapm}
    \gamma_1^\pm=\mp\frac{1}{2}(\gamma_2\pm\gamma_3), \ \gamma_2^\pm=\mp\frac{1}{2}(\gamma_3\pm\gamma_1), \ 
    \gamma_3^\pm=\mp\frac{1}{2}(\gamma_1\pm\gamma_2) \ .
\eeq

The $\gamma$-deformed theory \eqref{N=4SYMlagrangian} is non-supersymmetric but in the limit of equal deformations $\gamma_i=\beta$ a $\mathcal{N}=1$ supersymmetry is restored obtaining the Lunin-Maldacena setup \cite{Lunin:2005jy}. We will refer to the theory in this limit as $\beta$-deformed $\mathcal{N}=4$ SYM.
The action \eqref{N=4SYMlagrangian} was proposed as a non-supersymmetric example of AdS/CFT correspondence obtained by applying the deformation to both sides of the original duality \cite{Frolov:2005dj,Frolov:2005iq}. In particular, the string theory description is obtained acting with a TsT transformation\footnote{TsT stands for consecutive T-duality, shift and T-duality each depending on one of the parameters $\gamma_i$.} on the $S^5$ factor of the $AdS_5\times S^5$ background. 

In the last few years, a special limit of the $\gamma$-deformed theory that selects only simple chiral diagrams was proposed in \cite{Gurdogan:2015csr}. This is the limit of small 't Hooft coupling $g$ and large imaginary twists $\gamma_i$ such that their product $\xi_i=g\,e^{-i\gamma_i/2}$ is kept fixed. The resulting action is a non-unitary, non-supersymmetric CFT with three couplings $\xi_{1,2,3}$ known as the fishnet theory. The gauge fields and the gaugino decouple and one is left with three complex scalars and three complex fermions.
The interaction vertices impose a specific orientation on planar Feynman diagrams reflecting the chirality property of the theory. We will refer to this theory as $\chi CFT_4$. 
This theory was studied in \cite{Caetano:2016ydc} by the asymptotic Bethe ansatz methods. Furthermore, the spectrum of simplest operators were studied in \cite{Kazakov:2018gcy} together with exact results for certain four-point functions\footnote{In order to accomplish this goal, one uses the Bethe-Salpeter resummation of Feynman diagrams, conformal symmetry and the uniqueness method similarly to \cite{Grabner:2017pgm}. Moreover, the uniqueness method (known also as star triangle relations) was implemented in a Mathematica package in \cite{Preti:2018vog,Preti:2019rcq}.}.   

The $\chi CFT_4$ can be further reduced by tuning the value of the couplings. The simplest case is the single coupling reduction, $\xi_1=\xi_2=0$ and $\xi_3=\xi$, in which the action contains only two interacting complex scalar fields \cite{Gurdogan:2015csr}. This theory is known as bi-scalar fishnet theory. The planar Feynman graphs for typical physical quantities have a square fishnet structure where the massless scalar propagators form a regular quadratic lattice. The fishnet graphs appear to represent an integrable statistical mechanical system \cite{Zamolodchikov:1980mb}.

Similarly to the undeformed $\mathcal{N}=4$ SYM theory, also these deformations and reductions are most accessible in the ’t Hooft (planar) limit where the rank of the gauge group $SU(N_c)$ $N_c\rightarrow \infty$ and the coupling $g_{\rm YM}\rightarrow 0$ such that the ’t Hooft coupling $g=g_{\rm YM} \sqrt{N}/(4\pi)$ is kept fixed. In this limit, the string theory becomes free and in the gauge theory non-planar vacuum diagrams are suppressed.

\subsection{Integrability and BMN vacuum operators}

The $\gamma$-deformed theory shares certain properties with its parent $\mathcal{N}=4$ SYM theory. The most intriguing one is the claimed integrability in the planar limit. 
In \cite{Beisert:2005if} the one-loop dilatation operator was computed in terms of the original $\mathcal{N}=4$ SYM one. In addition to the deformed gravity background \cite{Frolov:2005dj}, this result is compatible with the AdS/CFT integrability \cite{Roiban:2003dw,Beisert:2005if}. 
Then, many other advanced integrability techniques were adapted to the study of the deformed theory such as the Y-system \cite{Gromov:2010dy}, TBA \cite{deLeeuw:2012hp} and finally the QSC \cite{Kazakov:2015efa}. The simplest test of the claimed integrability consists in the study of the spectrum of composite operators that are protected in the un-deformed $\mathcal{N}=4$ SYM theory but gain anomalous dimensions in the deformed case. 

Local operators in the theory \eqref{N=4SYMlagrangian} are classified with respect to the irreducible representations the global symmetry $SU(2,2)\otimes U(1)^{\otimes 3}$ and identified by the values of Cartan generators $(\Delta,S_1,S_2|J_1,J_2,J_3)$. Here $\Delta$ is the scaling dimension of the operator, the pair $(S_1,S_2)$ defines its Lorentz spin and $(J_1,J_2,J_3)$ the $U(1)_i$ charges.
In this paper, we will focus on the following simple class of operators
\beq\label{operators}
\mathcal{X}_J=\Tr{\left(\phi_1^J\right)},\qquad
\mathcal{Y}_J=\Tr{\left(\phi_2^J\right)},\qquad
\mathcal{Z}_J=\Tr{\left(\phi_3^J\right)}
\eeq
belonging to the representations $(\Delta_{\mathcal{X}_J},0,0|J,0,0)$, $(\Delta_{\mathcal{Y}_J},0,0|0,J,0)$ and $(\Delta_{\mathcal{Z}_J},0,0|0,0,J)$ respectively and with scaling dimension at zero coupling $\Delta=J$. Since such operators are determined uniquely by their global charges, operator mixing cannot occur. A similar class of local operators can be also defined in the parent $\mathcal{N}=4$ SYM theory. In that case, they are protected from quantum corrections and, since they correspond to ground states in the spin-chain picture, they are known as BMN vacuum operators. In the $\gamma$-deformed theory \eqref{N=4SYMlagrangian}, the operators \eqref{operators} are not protected anmymore\footnote{It is interesting to mention that those operators are still protected also in the $\beta$-deformed theory.} and their scaling dimensions start to depend on the coupling $g$ and the following combination of the twists
\beq\label{kappa}
   \kappa_j\coloneqq e^{i\,\gamma_j^+}\quad\text{and}\quad \hat\kappa_j \coloneqq e^{i\,\gamma_j^-}
   \qquad j=1,2,3\,,
\eeq
such that 
\beq\label{symdelta}
\Delta_{\mathcal{X}_J}=\Delta(g,\kappa_1,\hat\kappa_1)\,,\quad
\Delta_{\mathcal{Y}_J}=\Delta(g,\kappa_2,\hat\kappa_2)\,\quad\text{and}\quad
\Delta_{\mathcal{Z}_J}=\Delta(g,\kappa_3,\hat\kappa_3)
 \ .
\eeq
Since the function $\Delta$ is the same for the three scaling dimensions, in the following we will consider only the operator $\mathcal{X}_J$. In order to simplify the notation we also drop the indices on the twists such that $\gamma^\pm\coloneqq\gamma_1^\pm$, $\kappa\coloneqq\kappa_1$ and $\hat\kappa\coloneqq\hat\kappa_1$.

From the integrability point of view, in the $\gamma$-deformed theory the operators \eqref{operators} have an interesting feature. Indeed they do not receive corrections from the twisted Bethe ansatz at the asymptotic level, but solely from finite-size effects. The first two wrapping terms were computed in \cite{Ahn:2011xq} using the TBA and Y-system approach for $J\geq 3$. The case of $J=2$ has to be discussed separately. Indeed the equations of \cite{Ahn:2011xq} diverge and an integrability approach to this state has been missing. Moreover, quantum corrections induce the running of quartic scalar double-trace couplings \cite{Fokken:2013aea} as we discuss more closely in section \ref{sec:renormalization}.
For these reasons, in the following we will focus on this specific state studying the operator $\mathcal{X}_2$ by means of the twisted Quantum Spectral Curve, providing a crucial test of the integrability of the $\gamma$-deformed $\mathcal{N}=4$ SYM theory.

\subsection{Renormalization}\label{sec:renormalization}

An important feature of the theory \eqref{N=4SYMlagrangian} that is not rooted in the undeformed $\mathcal{N}=4$ SYM theory is the presence of scalar double trace counterterms. Those affect the anomalous dimension (and hence integrability) of the BMN operators in the $J=2$ case. Let's consider the operator $\mathcal{X}_2$ defined in \eqref{operators}. The double-trace counterterms to include in the Lagrangian \eqref{N=4SYMlagrangian} are the following
\beq\label{dt}
\mathcal{L}_{\text{dt}}=(4\pi)^2\,\alpha^2\,\Tr{\bigl(\phi_1^2\bigr)}\Tr{\bigl({\phi^{\dagger\,2}_1}\bigr)}+\dots \ ,
\eeq
where the first term renormalizes the operator we are considering and the dots represent all the other possible combinations of scalar double-traces that affect other operators. The coupling $\alpha=\alpha(g)$ runs with the 't Hooft coupling $g$ breaking the conformal invariance. The related $\beta$-function takes the following form \cite{Dymarsky:2005uh,Pomoni:2008de}
\beq\begin{split}\label{beta}
\beta_{\alpha^2}=&a(g) \alpha^4+b(g) \alpha^2+c(g)\\
=&4\alpha^4+g^4\frac{(\kappa^2-1)^2(\hat\kappa^2-1)^2}{\kappa^2 \hat\kappa^2}+\mathcal{O}(g^6)
\end{split} \ ,\eeq
where the second line was computed in \cite{Fokken:2013aea}. Imposing the vanishing of \eqref{beta}, we obtain the following fixed points at one-loop
\beq\label{fixedpoint}
\alpha^2_\pm=\pm \frac i2 g^2 \frac{(\kappa^2-1)(\hat\kappa^2-1)}{\kappa \hat\kappa}+\mathcal{O}(g^4) \ .
\eeq
Notice that the presence of the imaginary unit is a consequence of the fact that when the theory flows to the conformal points unitarity is broken.
The role of this double-trace counterterm in the $SU(N_c)$ theory can be interpreted in terms of the finite-size effect of pre-wrapping \cite{Fokken:2013aea,Fokken:2014soa}.
Indeed it contributes at the leading order in the large-$N_c$ expansion. This mechanism is similar to the wrapping one, but a contribution arises one order earlier in the perturbative expansion, namely the anomalous dimensions of a length-$J$ operator can start at order $g^{2J-2}$. In particular for the operators \eqref{operators}, this occurs only in the $J=2$ case we are studying. Indeed, the anomalous dimension of the operator $\mathcal{X}_2$ defined by $\Delta_{\pm}=2+\gamma_\pm$ takes the following form
\beq\label{gamma1}
\gamma_\pm=4\alpha^2_\mp-2g^4\frac{(\kappa^2-1)^2(\hat\kappa^2-1)^2}{\kappa^2 \hat\kappa^2}+\mathcal{O}(g^6) \ ,
\eeq
where the term proportional to $\alpha^2_\pm$ entirely originates from pre-wrapping. Furthermore, at the fixed points, it is possible to write this anomalous dimension in terms of the coefficients of \eqref{beta} as follows \cite{Pomoni:2008de}
\beq\label{gamma1}
4\gamma^2=b^2-4 a c \ .
\eeq
Using these relations, one can show that the RG-flow of the theory \eqref{N=4SYMlagrangian} supplemented by double-trace terms \eqref{dt} is defined solely in terms of the universal quantity $\gamma(g)$ \cite{Gromov:2018hut}. 
In section \ref{sec:weakres} we will derive the next-to-leading order term of the fixed points \eqref{fixedpoint} combining \eqref{gamma1} together with the results obtained from the solution of the twisted QSC. Similar arguments holds also for the fishnet cases \cite{Grabner:2017pgm,Gromov:2018hut,Kazakov:2018gcy,Pittelli:2019ceq,Pittelli:20191}.

\section{Quantum Spectral Curve for $\gamma$-deformed SYM}
\label{sec:qsc}

In this section we describe the Quantum Spectral Curve construction which is the basis of our results. We will be brief and refer the reader to  \cite{Gromov:2017blm,Kazakov:2018hrh,Levkovich-Maslyuk:2019awk} for recent reviews.

The QSC is a finite system of functional equations for a set of key objects known as Q-functions. It was originally formulated to describe the full spectrum of single-trace operators in planar $\cN=4$ SYM \cite{Gromov:2013pga}. Then it was generalized to the $\gamma$-deformed theory in \cite{Kazakov:2015efa}, with the only differences being in the large $u$ asymptotics of the Q-functions.

Among the most important Q-functions are
the 4+4 functions $\bP_a(u)$ and $\bP^a(u)$, $a=1,\dots,4$, which roughly speaking correspond to string motion on $S^5$. Their asymptotics encode the conserved angular momenta on $S^5$ as well as the corresponding twist angles $\gamma_i$. For our $J=2$ vacuum state they read
\beqa \label{asymptoticsPa}
&&\bP_a\sim  A_a x_a^{i u} u^{-\hat \lambda_a}\,,\qquad   \bP^a\sim  A^a x_a^{-i u} u^{\hat \lambda_a^*},
\eeqa
where
\beq
	x_a=\left\{\kappa^2,\kappa^{-2},\hat\kappa^2,\hat\kappa^{-2}\right\} \ ,
\eeq
\beqa\label{twists} \nn
&&\hat \lambda=\hat \lambda^*=\left\{1,1,-1,-1\right\} \ .
\eeqa
By using a rescaling symmetry we can set $A^a=1$ and then the remaining leading coefficients are given by
\beqa\label{AA}\nn
&&A_1=-A_2=\frac{\hat \kappa^2 (\kappa^2-1)^3}{(1+\kappa^2)(\kappa^2-\hat \kappa^2)((\kappa \hat \kappa)^2-1)}\\
&&A_3=-A_4=-\frac{\kappa^2 (\hat\kappa^2-1)^3}{(1+\hat\kappa^2)(\kappa^2-\hat\kappa^2)((\kappa \hat \kappa)^2-1)}\;.
\eeqa
The $\bP$-functions can be parameterized concisely in terms of a set of coefficients $c_{a,n}$ that are the main parameters encoding all the nontrivial data about the state and conserved charges. The parameterization was worked out in \cite{Gromov:2017cja,Kazakov:2015efa} and for our case reads
\beqa
\label{pasgen}
\bP_a(u)=x_a^{i u} (g x(u))^{-\hat \lambda_a}\bp_a(u)\;\;,\;\;
\bP^a(u)=x_a^{-i u} (g x(u))^{\hat \lambda_a^*}\bp^a(u),
\eeqa
where we introduced
\beqa\label{Pa1}\nn
\label{Pa2}\nn
&&\bp_a=\left\{A_1 f_1(u),A_2 f_1(-u),A_3 g_1(u),A_4 g_1(-u)\right\} \ ,\\[2mm]
&&\bp^a=\left\{ f_2(u), f_2(-u), g_2(u), g_2(-u)\right\} \ ,
\eeqa
and the functions $f_1,f_2,g_1,g_2$ are series of the form
\beqa\label{fg_ansatz}
&&f_1=1+g^{4} \sum\limits_{n=1}^{\infty}\frac{g^{2n-2}c_{1,n}}{(g x)^n}\\
&&g_1=(g x)^{-2}\left(u^2+  \sum_{k=0,1} c_{2,-k}u^k+ \sum\limits_{n=1}^{\infty}\frac{g^{2n}c_{2,n}}{(g x)^n}\right)\\
&&f_2=(g x)^{-2}\left(u^2+  \sum_{k=0,1} c_{3,-k}u^k+ \sum\limits_{n=1}^{\infty}\frac{g^{2n}c_{3,n}}{(g x)^n}\right)\\
&&g_2=1+g^{4} \sum\limits_{n=1}^{\infty}\frac{g^{2n-2}c_{4,n}}{(g x)^n} \ .
\eeqa
Here we use the standard Zhukovsky variable $x(u)$ defined by
\beq
    x+\frac{1}{x}=\frac{u}{g} \ , \ \ |x|>1 \ .
\eeq
The extra factors of $g$ appearing in this parameterization ensure that at weak coupling $c_{a,n}\sim 1$ (as we will also see explicitly from the solution of the QSC).

The AdS$_5$ counterpart of the $\bP$-functions are the $4+4$ functions $\bQ_i(u)$ and $\bQ^i(u)$, $i=1,\dots,4$. Their asymptotics encode the AdS conserved charges including $\Delta$, which for our state amounts to simply
\beqa  \label{qasgen}
&&\bQ_i\sim B_i u^{-\hat \nu_i}\(1+\sum_{k=1}^\infty\frac{B_{i,k}}{u^{2k}}\) \ , \ \ 
\bQ^i\sim B^i u^{\hat \nu_i^*}\(1+\sum_{k=1}^\infty\frac{B^{i,k}}{u^{2k}}\)
 \ ,
\eeqa
with
\beqa\label{qas}
&& {\mathbf -}\hat\nu_i=\left\{\frac{\Delta}{2},1+\frac{\Delta}{2},2-\frac{\Delta}{2},3-\frac{\Delta}{2}\right\},\\
&&\hat \nu_i^*=\left\{-\frac{\Delta}{2}+3,-\frac{\Delta}{2}+2,\frac{\Delta}{2}+1,\frac{\Delta}{2}\right\}\,.
\eeqa
Note that the large $u$ expansion of the $\bQ$-functions goes in even powers of $u$ for our case.
We have
\beqa
&&B^1B_1=-B^4B_4 =\frac{i(\kappa^2-1)^2(\hat \kappa^2-1)^2}{(\kappa \hat \kappa)^2(\Delta-2)(\Delta-3)}\notag \ ,\\
&&B^2B_2=-B^3B_3=-\frac{i(\kappa^2-1)^2(\hat \kappa^2-1)^2}{(\kappa \hat \kappa)^2(\Delta-1)(\Delta-2)}
\label{BB}\;.
\eeqa
The $\bQ$-functions are indirectly fixed in terms of $\bP$'s as the solutions to the 4th order Baxyter-type equation
(first described in \cite{Alfimov:2014bwa}) which has the form
\beqa\nn
\bQ^{[+4]}_iD_0
&-&
\bQ_i^{[+2]}
\[
D_1-\bP_a^{[+2]}\bP^{a[+4]}D_0
\]
+
\bQ_i
\[
D_2-\bP_a\bP^{a[+2]}D_1+
\bP_a\bP^{a[+4]}D_0
\]\\
&-&
\bQ_i^{[-2]}
\[
\bar D_1+\bP_a^{[-2]}\bP^{a[-4]}\bar D_0
\]
+\bQ_i^{[-4]}\bar D_0
=0\la{bax5} \ ,
\eeqa
where $D_k$ are some determinants built from $\bP$'s which we give in appendix~\ref{app:qsc}, and we used the notation
\beq
    f^\pm=f(u\pm i/2) \ , \ \ \ f^{[+a]}=f(u+ia/2) \ .
\eeq
The $\bQ^i$ functions satisfy a similar equation. Let us also mention that we have
\beq
\label{PPz}
    \bP_a\bP^a=0 \ , \ \ \ \bQ_i\bQ^i=0 \ .
\eeq

While the $\bP$-functions are analytic except for one branch cut at $u\in[-2g,2g]$, the $\bQ$'s have an infinite set of cuts at $u\in[-2g+in,2g+in]$, $n=0,1,2,\dots$. To fix the solution of the QSC it remains to impose gluing conditions that relate $\bQ$'s and their analytic continuation around the branch point at $u=2g$, which we denote by $\tilde \bQ$. In our case it follows from the discussion in \cite{Gromov:2015vua,Gromov:2017cja} that the gluing conditions read
\beq
\label{gluing}
    \tilde \bQ_1(u)=\alpha_1\bQ^2(-u), \ 
    \tilde \bQ_2(u)=\alpha_2\bQ^1(-u), \ 
    \tilde \bQ_3(u)=\alpha_3\bQ^4(-u), \ 
    \tilde \bQ_4(u)=\alpha_4\bQ^3(-u) \ ,
\eeq
where $\alpha_i$ are some constants. Here we have used that the solution for the ground state should respect the $u\to -u$ symmetry. These conditions fix all the coefficients $c_{a,n}$ and most importantly the scaling dimension $\Delta$.

\subsection{Asymptotics and symmetries}

Let us describe some additional technical but important points of the QSC formulation in our case.

First, the large $u$ asymptotics of the $\bP$-functions described in \cite{Kazakov:2015efa} contain only information about the form of the asymptotics \eq{asymptoticsPa} and the values of the leading coefficients $A_i,A^i$. However, this is not sufficient as we should ensure that subleading coefficients satisfy a set of constraints which guarantee that the Baxter equation \eq{bax5} gives $\bQ$-functions with the prescribed asymptotics \eq{qasgen}, \eq{qas}. To derive these constraints, we plug the large $u$ expansion of $\bP$'s into the Baxter equation \eq{bax5} and deduce asymptotics of the solution. This computation is quite nontrivial and one has to expand the equation to a rather high order in $1/u$, essentially because the large $u$ asymptotics of $\bP$'s in \eq{pasgen} contains 4 distinct exponential twists while the $\bQ$-functions are not twisted at all. As a result, we find a set of 6 nontrivial constraints for the first few $c_{a,n}$ coefficients in the $\bP$-functions\footnote{Some of these constraints were previously derived by N. Gromov, V. Kazakov and G. Sizov whom we thank for sharing their results.}. The first two of them read
\beqa
\label{cas1}
    c_{1,1}&=&-\frac{c_{3,-1}}{g^4}-\frac{2 i \kappa ^2 \left(\kappa ^2+1\right) \left(\hat{\kappa
   }-1\right)^2 \left(\hat{\kappa }+1\right)^2}{g^4 (\kappa -1) (\kappa +1) \left(\kappa
   -\hat{\kappa }\right) \left(\kappa +\hat{\kappa }\right) \left(\kappa  \hat{\kappa }-1\right)
   \left(\kappa  \hat{\kappa }+1\right)} \ , \\
   \label{cas2}
   c_{4,1}&=& -\frac{c_{2,-1}}{g^4}+\frac{2 i (\kappa -1)^2 (\kappa +1)^2 \hat{\kappa }^2 \left(\hat{\kappa
   }^2+1\right)}{g^4\left(\hat{\kappa }-1\right) \left(\hat{\kappa }+1\right) \left(\hat{\kappa
   }-\kappa \right) \left(\kappa +\hat{\kappa }\right) \left(\kappa  \hat{\kappa }-1\right)
   \left(\kappa  \hat{\kappa }+1\right)} \ .
\eeqa
We give the full set of these relations in a Mathematica notebook accompanying this paper, as the remaining ones are rather lengthy. In addition, there is a constraint relating $c_{a,n}$ with $\Delta$ \cite{Gromov:2017cja} which we give in appendix \ref{app:qsc}. Let us also point out that the $c_{a,n}$ are further constrained by the relation $\bP^a\bP_a=0$.

Another important complication is that our system does not have an immediate left-right symmetry, i.e. the Q-functions with upper and lower indices are not related in a trivial way. This is in contrast to simple examples like $sl(2)$ sector in the undeformed model where they differ by just a sign and relabelling of indices. Nevertheless, we do have a version of the symmetry, where we also need to exchange the twists $\kappa \leftrightarrow \hat\kappa$ when we raise the indices, that is
\beq
\label{upsym}
    \bP^a(u)=r^a\sum_{b=1}^4\chi^{ab}\bP_b(u)\big|_{\kappa\leftrightarrow\hat\kappa} \ , \ 
    \bQ^i(u)=s^i\sum_{j=1}^4\chi^{ij}\bQ_j(u)\big|_{\kappa\leftrightarrow\hat\kappa} \ . \ 
\eeq
Here $r^a$ and $s^i$ are some constants depending on the way the rescaling symmetry for the $\bP$- and $\bQ$-functions is fixed, while $\chi^{ab}$ is the matrix
\beq\label{chi}
\chi= {\small  \left( \begin{matrix}
                0&0&0&-1 \\
                0&0&1&0 \\
                0&-1&0&0 \\
                1&0&0&0
        \end{matrix}\right)} \ .
\eeq
Relations of this kind were pointed out in \cite{Kazakov:2015efa,Gromov:2017cja}. Importantly, the relations \eq{upsym} for the $\bP$-functions imply that in our parameterization \eq{pasgen}-\eq{fg_ansatz} we have
\beq
\label{csymm}
	c_{1,n}=(-1)^{n}c_{4,n}\big|_{\kappa\leftrightarrow\hat\kappa} \ , \ c_{2,n}=(-1)^{n}c_{3,n}\big|_{\kappa\leftrightarrow\hat\kappa} \ .
\eeq
These equations relate $c$'s evaluated at different values of the twists, and as such they are not useful in practice for the numerical solution where we are solving the system at a fixed value of all the parameters. However, they lead to nice simplifications in the analytic perturbative solution of the QSC where we find $c_{a,n}$ as explicit functions of the twists at each order in $g$. We will give some more examples and details in the next section. Let us also note that the relations \eq{cas1} and \eq{cas2} we just discussed are compatible with the symmetry \eq{csymm}. 

%fix the freedom of rescaling the $\bP$- and $\bQ$-functions. So far only the combinations $A_aA^a$ and $B_iB^i$ were fixed by \eq{}, \eq{} (no summation over repeated indices here), and we can still use the rescaling transformation
%\beq
%    \bP_a\to r_a \bP_a \ , \ \ \ \bP^a\to 1/r_a \bP^a
%\eeq
%paremeterized by 4+4 constants 

\section{Weak coupling solution and results}

\label{sec:sols}

In this section we will describe the weak coupling perturbative solution of the QSC. We use the standard algorithm of solving the QSC iteratively \cite{Gromov:2015vua} (and the Mathematica package accompanying the paper \cite{Gromov:2015dfa}). We briefly summarize it below, highlighting the special features of the case we consider.

Since we are looking at the operator with
\beq
    \Delta=2+\cO(g^2) \ ,
\eeq
one may suspect a potential problem due to the asymptotics of the $\bQ$-functions \eq{BB} which contain a $1/(\Delta-2)$ factor that becomes singular at weak coupling. This difficulty was also pointed out in \cite{Kazakov:2015efa}. However, it merely means that some of the Q-functions are singular for small $g$, and does not lead to any singularity in the scaling dimension $\Delta$ itself as we will soon see. In fact, the QSC has already been successfully used for a setup with similar singular asymptotics, e.g. in \cite{Gromov:2015dfa,Gromov:2016rrp}.

To begin, we write all the coefficients $c_{a,n}$ as a power series in the coupling\footnote{Strictly speaking, the fact that the expansion goes in even powers may be viewed as an assumption we make. It is strongly supported by the agreement between our perturbative solution and numerical solution we discuss below, and also by the fact that we reproduce analytically many known results at weak coupling.}
\beq\label{ccoeff}
    c_{a,n}=\sum_{k=0}^\infty c_{a,n,k}g^{2k} \ ,
\eeq
and we assume that they all start from the $g^0$ term, i.e. are non-singular, which is also confirmed by the numerical solution at finite coupling we present below. Already at this stage we can fix some of the $c_{a,n,k}$ by imposing the constraints of the type \eq{cas1}, \eq{cas2} following from the large $u$ expansion\footnote{Namely, requiring that all $c_{a,n}$ are nonsingular for $g\to 0$ allows us to fix some of the $c_{a,n,k}$ coefficients.}. Next we expand the $\bP$-functions given by \eq{pasgen}-\eq{fg_ansatz} for small $g$ and plug them into the Baxter equation \eq{bax5}. Our first goal is to generate a basis of 4 solutions of this equation to a high order in the coupling. At leading order we can guess the solutions which read
\beq
\label{qg0}
\begin{split}
    q_I^{(0)}&= u^2 \,, \qquad q_{II}^{(0)} = -u^2 \,\eta_2(u)+u\, \eta_1(u)-\frac{i}{u}\,\frac{\kappa^2}{(\kappa^2-1)^2} \ ,\\
    q_{III}^{(0)} &= u \,, \qquad\;\, q_{IV}^{(0)} =-u^2\, \eta_1(u)-i\frac{\kappa^2}{(\kappa^2-1)^2} \ .
\end{split}
\eeq
Here we introduced the usual $\eta$-functions defined by
\beq
\eta_{s_1,\dots,s_k}(u)=\sum_{n_1>n_2>\dots>n_k \geq 0}
\frac{1}{(u+in_1)^{s_1}(u+in_2)^{s_2}\dots (u+in_k)^{s_k}} \ ,
\eeq
which typically appear in the weak coupling solution of the QSC \cite{Marboe:2014gma,Leurent:2013mr}.  Next we increase the accuracy of these solutions order by order in $g$. To do this in practice, we introduce the $Q_{a|i}$ Q-functions as solutions of the equation
\beq
	Q_{a|i}^+-Q_{a|i}^-=\bP_a\bQ_i \ .
\eeq
Instead of the 4th order Baxter equation, we can solve a system of 1st order equations on these functions,
\beq
\label{QPPQ}
  Q_{a|i}^+-Q_{a|i}^-+\bP_a\bP^bQ_{b|i}^+=0
\eeq
and then reconstruct $\bQ_i$ due to the property 
\beq
\label{QPQ}
	\bQ_i=-\bP^aQ_{a|i}^+ \ .
\eeq
We can easily translate our starting solutions \eq{qg0} to a basis of solutions of \eq{QPPQ} at leading order in the coupling. Then using the iterative method of \cite{Gromov:2015vua} we solve the equation \eq{QPPQ} order by order in the coupling, and lastly use \eq{QPQ} to find a basis of four solutions $q_I, q_{II}, \dots q_{IV}$ to the Baxter equation for $\bQ_i$. Finally, we construct the true $\bQ_i$ functions as linear combinations of these four solutions, fixed by imposing the large $u$ behavior \eq{qasgen}. 

At this point we encounter an important technical difficulty. In order to proceed with imposing the gluing conditions we also need to know the $\bQ^i$ functions in addition to $\bQ_i$. In principle they can be found by constructing $Q^{a|i}$ with upper indices defined as minus the inverse transposed matrix\footnote{They also satisfy an equation similar to \eq{QPPQ} which reads $	Q^{a|i+}-Q^{a|i-}=\bP^a\bQ^i$.} to $Q_{a|i}$,
\beq
  Q_{a|i}Q^{a|j}=-\delta_i^j \ ,
\eeq
and then contracting it with the $\bP$-functions to get
\beq
\label{Qiup}
	\bQ^i=+\bP_aQ^{a|i+} \ .
\eeq
However, it is highly time consuming to actually invert the complicated $Q_{a|i}$ matrix. Luckily, instead of this we can make a shortcut by invoking the symmetry \eq{upsym} which means that\footnote{up to an constant overall factor which is not important for us due to the linear form of the gluing conditions \eq{gluing}.} $\bQ^i$ is given by $\bQ_i$ albeit with twists $\kappa$ and $\hat\kappa$ exchanged. At this point we have $\bQ_i$ written in terms of the $c_{a,n,k}$ coefficients, and nicely we know that exchanging the twists in these coefficients simply amounts to relabeling them according to the rules \eq{csymm} (we remind that the $c_{a,n,k}$ are the terms in the weak coupling expansion of the $c_{a,n}$ coefficients). As a result, we get the $\bQ^i$ functions almost for free! 

Having found both $\bQ_i$ and $\bQ^i$ in terms of the $c_{a,n,k}$ coefficients, we finally impose the gluing conditions \eq{gluing} order by order in $g$. This fixes all the unknowns, and provides the result for the scaling dimension $\Delta$.

Let us also note that due to tricky cancellations at intermediate steps as well as the $1/(\Delta-2)$ singularity in the asymptotics \eq{BB}, we actually have to compute the Q-functions to several orders higher in $g$ than the order at whcih we wish to fix the scaling dimension. E.g. to get even the 1-loop term in $\Delta$ we already need several orders in the expansion of the Q's (a similar thing happens for instance for the quark-antiquark potential \cite{Gromov:2016rrp}). 

As a consistency check, we also performed the first few orders of the computation separately without using the twist exchange symmetry \eq{upsym}, \eq{csymm} at all, and rather inverting the matrix $Q_{a|i}$ directly and then computing $\bQ^i$ from \eq{Qiup}. We verified that in this way we obtain the same scaling dimension at least to the first couple of orders in $g$. 

Let us also mention that in practice it was often useful for us to utilize not only the gluing conditions, but also the equation
\beq\label{Qtilde}
\tilde\bQ_i=-\tilde\bP^aQ_{a|i}^+
\eeq
(a consequence of \eq{QPQ} and the fact that $Q_{a|i}$ has no cuts in the upper half-plane), which sometimes allows us to fix some unknown coefficients without going to an unncessarily high order in the small $g$ expansion. This is just a technical trick, and we have verified numerically that the gluing conditions alone are sufficient to fully fix the solution (see section~\ref{sec:num}).

\subsection{Weak coupling results}\label{sec:weakres}

Using the method described above, we have solved the QSC analytically to a high order in the weak coupling expansion. We have generated the functions $\bQ_i$ and $\bQ^i$, then imposing the gluing conditions\footnote{Using equation \eqref{Qtilde}, this is equivalent to impose that $\bQ_i+\tilde \bQ_i$ and $(\bQ_i-\tilde \bQ_i)/\sqrt{u^2-4g^2}$ are regular for $u\rightarrow 0$} we fix the $c_{a,n,k}$ coefficients and the scaling dimension.

Let us note that as a result we find two solutions for the scaling dimension, related by $\Delta \leftrightarrow 4-\Delta$. This is expected from the renormalization group arguments presented in section \ref{sec:renormalization}. Indeed, the two solutions corresponds to choice of one of the two fixed points \eqref{fixedpoint} as explicitly shown in the relation \eqref{gammapm}. Moreover, both of the solutions correspond to the same values of the $c_{a,n}$ coefficients. This also is expected immediately because of the relation between $\Delta$ and $c_{a,n}$ given in \eq{eqDelta}. One solution can be interpreted as a "\textit{physical state}" while the other is a "\textit{shadow state}". Indeed, given a physical operator with dimension $\Delta_{ph.}$, in 4d the shadow operator has scaling dimension given by $\Delta_{sh.}=4-\Delta_{ph.}$. In our case, both of them have real part equal to 2, and opposite imaginary parts. This is compatible with the fact that the operator $\mathcal{X}_2$ we are studying is with no spin, then the scaling dimensions of the physical and shadow states are simply related by complex conjugation. We label these solutions as $\Delta_\pm$ according to the sign of the imaginary part. 

As an example, the $\bQ_i$ functions read, to $g^2$ accuracy (up to an overall normalization)
\beq\begin{split}
\bQ_{1(3)}&=u\pm g^2\left[2S_-S_+u[2\left(\eta_1-i-u\eta_2\right)-\pi]+i\frac{ S_- \mp i S_+}{u S_+}\right]+\dots \ ,\\
\bQ_{2(4)}&=u^2\pm g^2\left[2S_-S_+u[(2\eta_1-\pi) u-1]-i\frac{S_- \mp i S_+}{S_+}\right]+\dots \ ,
\end{split}\eeq
where we used the shorthand notation
\beq\label{spsm}
    S_+=-\frac{i}{2}\frac{\kappa^2-1}{\kappa}=\sin\gamma_1^+\,,\qquad\text{and}\qquad
    S_-=-\frac{i}{2}\frac{\hat\kappa^2-1}{\hat\kappa}=\sin\gamma_1^-\,.
\eeq
Let us also note that the twists exchange $\kappa\leftrightarrow \hat\kappa$ can be easily translated in this notation as $S_-\leftrightarrow S_+$. As another example, one of the $c_{a,n}$ coefficients reads to order $g^{10}$:
\beq\begin{split}
c_{3,-1}=2S_-^2S_+\sqrt{1-S_+^2}\biggl[
&\frac{1}{S_+^2(S_-^2-S_+^2)}
+32 \zeta_3 g^6
-560\zeta_5g^8\\
&-128(4S_-^2S_+^2(\zeta_3+\zeta_5)+6S_+^2(\zeta_3)^2-63\zeta_7)g^{10}+\mathcal{O}(g^{12})
\biggr]
\end{split}\eeq
and $c_{2,-1}=-c_{3,-1}\big|_{S_- \leftrightarrow S_+}$.

As a final outcome, we obtain the five-loop weak coupling expansion of the scaling dimension, which was advertised in \eq{Dintro} in the introduction and is one of our main results. Since we would like to discuss some of its features, for reading convenience we repeat it here:
\beqa\label{weakcoupdelta}
    \Delta_\pm=2 &\pm& 8 i S_- S_+{\color{blue}g^2}
    \\ \nn &\pm&
     0\times {\color{blue}g^4}
     \\ \nn &\pm&
     32iS_-S_+\[-3   \left(S_-^2+S_+^2\right)  \zeta_3-2  S_-^2 S_+^2\]{\color{blue}g^6}
     \\ \nn
    &\pm&256iS_-S_+\[
     4  S_-^2 S_+^2 \zeta_3+5   \left(S_-^2+S_+^2\right)  \zeta_5 \]{\color{blue}g^8}
    \\ \nn
    &\pm&64iS_-S_+\[28  S_-^4 S_+^4+36  S_-^2 S_+^2 \left(S_-^2+S_+^2\right)
   \zeta_3+20  S_-^2 S_+^2\left(3 S_-^2+3 S_+^2-13\right) \zeta_5
      \right. \\ \nn && \left.
   +9  \left(3 S_-^4-14
   S_+^2 S_-^2+3 S_+^4\right) (\zeta_3)^2 -245   \left(S_-^2+S_+^2\right)
   \zeta_7\]{\color{blue}g^{10}}+{\cal O}({\color{blue}g^{12}}) \ .
\eeqa
Notice that the dimensions \eqref{weakcoupdelta} are symmetric in the exchange $S_- \leftrightarrow S_+$. This feature is expected, and it follows from the QSC description as can be easily verified using \eqref{eqDelta} and \eqref{csymm}.

%The first non-trivial check for our formula is that it reproduces the correct scaling dimension in the $\beta$-deformed case. This corresponds to the limit of equal twists such that $S_+=\sin\beta$ and $S_-=0$. Then, since any order in \eqref{weakcoupdelta} is proportional to $S_-$, we conclude that the $\mathcal{X}_2$ operator is protected in the $\beta$-deformed theory where a $\mathcal{N}=1$ supersymmetry is restored. Moreover, we also exactly reproduce the $g^2$ order predicted in  \cite{Fokken:2014soa}. 

The first non-trivial check of our result is that we exactly reproduce the $g^2$ order predicted in  \cite{Fokken:2014soa}! Moreover, the fact that the order $g^4$ vanishes has interesting diagrammatic consequences. Feynman diagrams appearing at this order are the one of the previous order dressed by a gluon propagator or with propagators substituted by their one-loop self-energies. In dimensional regularisation ($D=4-2\epsilon$), from the RG equations it is possible to show that at the fixed points \eqref{fixedpoint} there are no divergencies higher than $1/\epsilon$ \cite{Fokken:2014soa}. Since the order $g^4$ vanishes, we expect that also the $1/\epsilon$ divergence disappears. It could be interesting to verify this statement with a diagrammatic approach. Finally, combining the result $\Delta_\pm$ together with \eqref{gammapm}, it is possible to compute the fixed point \eqref{fixedpoint} up to order $g^4$ obtaining
\beq
\alpha^2_\pm=\mp \,2 i \,S_+S_- \,g^2
+8\, S_+^2S_-^2 \,g^4
+\mathcal{O}(g^6) \ .
\eeq
The leading order matches the result of \cite{Fokken:2014soa}. Unfortunately we do not have enough relations to fix the $\beta$-function to higher orders. Indeed using the first line of \eqref{beta} and the relation \eqref{gamma1} at the fixed point, it is possible to write two of the three coefficients of the $\beta$-function, for instance $a$ and $b$, in terms of the dimension \eqref{weakcoupdelta} and the remaining coefficient, for instance $c$. Then the computation of the $\beta$-function can be reduced to the computation of the $1/\epsilon$ divergencies of diagrams with only single-trace vertices. We leave this point for the future.  

Let us also discuss the cases when the $\gamma_i$ parameters take special values. We can immediately see that for the $\beta$-deformed theory (with all $\gamma_i=\beta$) we have $S_-=0$, so all terms in our result vanish giving $\Delta=2$, as expected since this state becomes protected by $\cN=1$ supersymmetry. Another special case is a partial $\gamma$-deformation, for instance when  $\gamma_1=\gamma_2=0$. Then $S_+=S_-=-\sin\frac{\gamma_3}{2}$ and the result slightly simplifies,
\beqa\label{weakcoupdelta2}
\begin{split}
    \Delta_\pm&\big|_{S_+=S_-=S}=2 
    \pm 8 i S^2{g^2}
    \mp 64iS^4(S^2+3S^2\zeta_3){g^6}
    \pm 512 i S^4(2S^2\zeta_3+5\zeta_5){g^8}\\
    &\pm 128 i S^4(2S^2(7S^4+6S^2(3\zeta_3+5\zeta_5)-65\zeta_5-18(\zeta_3)^2-245\zeta_7)){g^{10}}
     +{\cal O}({g^{12}}) \ .
     \end{split}
\eeqa

\subsection{Comparison with fishnet theories}

In this section we compare our five-loop result for the scaling dimension of the operator $\mathcal{X}_2$ \eqref{weakcoupdelta} with the known results for the same operator in fishnet theories. In order to do this, we will study the expansion \eqref{weakcoupdelta} in the double scaling (DS) limit of small 't Hooft coupling $g$ and large imaginary twists $\gamma_i$ such that their product $\xi_i=g\,e^{-i\gamma_i/2}$ is kept fixed. Then we can reshuffle the expansion \eqref{weakcoupdelta} as follows 
\beq\label{exp}
\Delta_\pm=
\Delta_{\pm}^{DS-LO}+
\Delta_{\pm}^{DS-NLO}\frac{1}{\kappa}+
\Delta_{\pm}^{DS-NNLO}\frac{1}{\kappa^2}+\mathcal{O}\left(\frac{1}{\kappa^3}\right) \ ,
\eeq
where the term $1/\kappa$ plays the same role of $1/N_c$ corrections in the 't Hooft limit. Then, the leading order term in the DS-limit is given by
\beq\label{DSLO}
\Delta_{\pm}^{DS-LO}=
2
\pm 2i\omega^2
\mp i\omega^2[\omega^4\!\!-6\rho^4\zeta_3]\pm \frac{i}{4}\omega^2[7\omega^8-12\omega^4\rho^4(3\zeta_3+5\zeta_5)+108\rho^8\zeta_3^2]+\dots \ ,
\eeq
where we defined the following combinations of the double-scaled couplings
\beq
\omega^4=(\xi_2^2-\xi_3^2)^2 \ ,\qquad \ 
\rho^4=\xi_2^2\xi_3^2 \ ,
\eeq
and dots represent higher orders in $\omega$ and $\rho$.
The expansion \eqref{DSLO} exactly matches the weak coupling scaling dimension of the same operator in $\chi CFT_4$ given in \cite{Kazakov:2018gcy} ! This serves as a nontrivial check of our QSC calculation. The exact result of \cite{Kazakov:2018gcy} for the spectrum can be easily expanded at weak coupling, and in general is given in terms of the 
%of $\mathcal{X}_2$ was computed exactly in \cite{Kazakov:2018gcy} in terms of the
following implicit equation\footnote{We use a different notation respect to \cite{Kazakov:2018gcy} where $\xi_2^2\xi_3^2=\kappa^4$.}
\begin{align}\label{speceq}
\frac{\Delta(\Delta\!-\!2)^2(\Delta\!-\!4)}{16}\biggl[\!1+&\frac{\rho^4}{2-\Delta}\!\left[ \psi ^{(1)}\!\left(1\!-\!\tfrac{\Delta}{4} \right)
\!-\!\psi ^{(1)}\!\left(\tfrac{3}{2}\!-\!\tfrac{\Delta}{4} \right)\!+\!
\psi ^{(1)}\!\left(\tfrac{1}{2}\!+\!\tfrac{\Delta}{4} \right)\!-\!
\psi ^{(1)}\!\left(\tfrac{\Delta}{4}\right)\right]\!\!
\biggl]=\omega^4,
\end{align}
where $\psi^{(1)}(x)=d\psi(x)/dx$ and $\psi(x)$ is the digamma function. Given the ansatz $\Delta=2+\sum_{k,j>0} \alpha_{2k,2j}\, \omega^{2k}\rho^{2j}$ for the scaling dimension, it is possible to solve perturbatively the equation \eqref{speceq} up to any desired order in the couplings.

Furthermore, we can explore the sub-leading orders of our result in the DS-limit.
Considering that $\omega,\rho\sim g$, the next-to-leading starts at order $g^8$ and it is given by
\beq
\Delta_{\pm}^{\rm DS-NLO}=
\pm 16 i \rho^2\omega^2[\omega^4\zeta_3-5\rho^4\zeta_5]+\dots \ ,
\eeq
and the NNLO starts as $g^2$ and it is given by
\beqa
&&\Delta_{\pm}^{\rm DS-NNLO}=
\mp 2i\omega^2
\pm 3i\omega^2[\omega^4(2\zeta_3+1)-6\rho^4\zeta_3]\\
&&\mp \frac{i}{4}\omega^2[\omega^8(36\zeta_3+60\zeta_5+35)+4\omega^4\rho^4(9\zeta_3(14\zeta_3-5)+185\zeta_5)+20\rho^8(27\zeta_3^2-196\zeta_7)]+\dots\nonumber \ .
\eeqa
Analysing the following subleading orders and considering the exact result of \cite{Kazakov:2018gcy}, we can formulate the following guess. The coefficients in \eqref{exp} of even powers of $1/\kappa$ are series in $g^{4n+2}$ with $n=0,1,2,\dots$ and the coefficients of odd powers of $1/\kappa$ are series in $g^{4n+4}$.

Finally, we can also compare with the results for the simplest bi-scalar fishnet theory \cite{Gurdogan:2015csr} setting $\rho=0$ and $\omega=\xi$. We obtain the following expansion.
\beq\begin{split}\label{bis}
\Delta_\pm^{\rm biscalar}=&[2\pm 2i\xi^2\mp i\xi^6\pm\frac 74 i\xi^{10}+\dots]\\
+&[\mp2i\xi^2\pm i\xi^6(2\zeta_3+1)\mp\frac i4 \xi^{10}(36\zeta_3+60\zeta_5+35)+\dots]\frac{1}{\kappa^2}+\mathcal{O}\left(\frac{1}{\kappa^4}\right) \ .
\end{split}\eeq
The first line contains the first few terms of the anomalous dimension of the operator $\mathcal{X}_2$ in the bi-scalar theory \cite{Grabner:2017pgm}. It is given to all orders by the  following simple formula
\beq
\Delta_{\pm}=2\pm\sqrt{2-2\sqrt{1+4\xi^4}} \ ,
\eeq
which can also be found by solving the equation \eqref{speceq} in the case of $\rho=0$ and $\omega=\xi$.
Expanding it at weak coupling $\xi$ we match the LO of \eqref{bis}.
The second line in \eqref{bis} is the NLO prediction in the DS-limit following from our result. Notice that the subleading orders in the DS-limit are going in powers of $1/\kappa^2$. Indeed, the coefficients of odd powers of $1/\kappa$ in \eqref{exp} are always proportional to some positive power of $\rho$ that vanishes in this case.

\section{Numerical results}
\label{sec:num}

In addition to the analytic results described above, we have solved the QSC for the $J=2$ state in a wide range of the coupling numerically with high precision, following the efficient algorithm developed in \cite{Gromov:2015wca}. One complication in our case is the absence of symmetry between Q-functions with upper and lower indices. The twist exchange relations \eq{upsym} do not seem useful for numerics as we are working at a fixed value of the twists, but it is straightforward to implement the algorithm regardless of this \footnote{see also \cite{Alfimov:2018cms} for a numerical solution of the QSC without left-right symmetry}. The lack of symmetry, however, increases the computation complexity and time. Another complication is the need for a good starting point, as the relations of the type \eq{cas1}, \eq{cas2} between the $\bP$-function coefficients make them singular at weak coupling unless one carefully chooses initial values. In practice we used the results from the perturbative solution as an input for the numerical algorithm at weak coupling.

\begin{figure}
    \centering
    \includegraphics[scale=1.0]{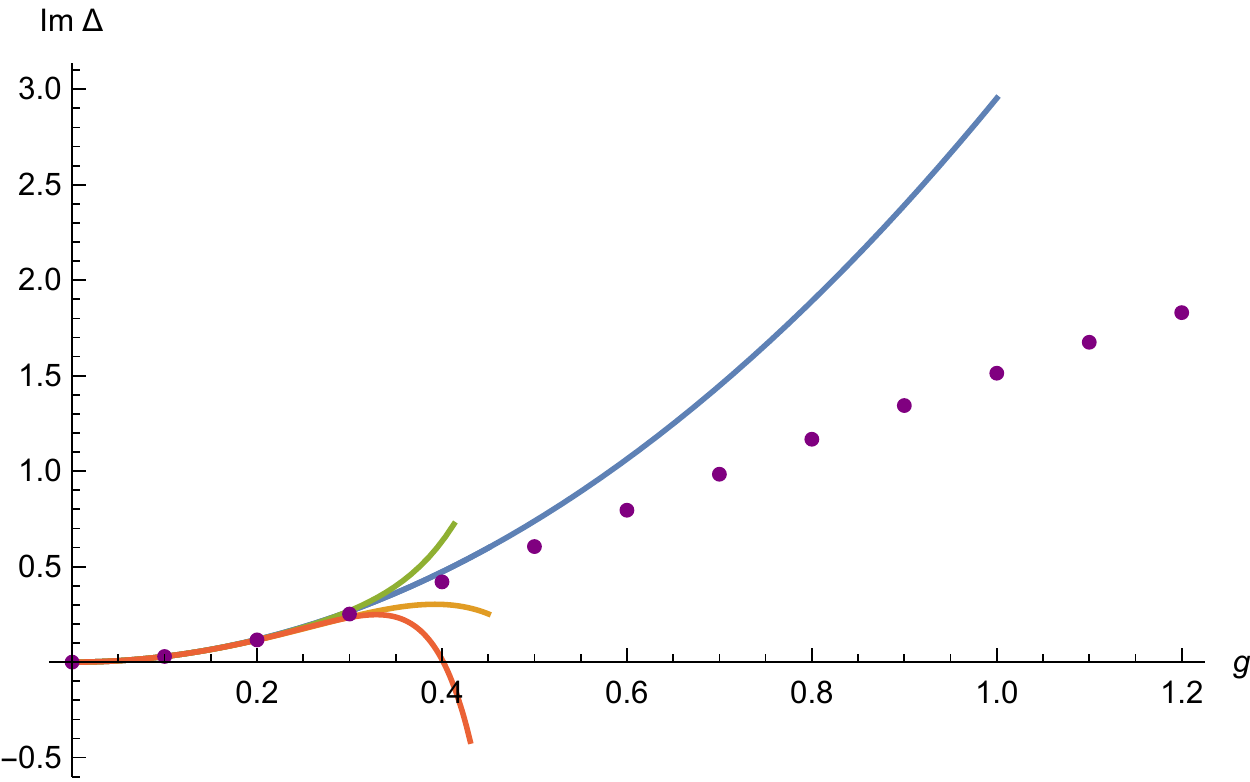}
    \caption{Numerical results for the scaling dimension of the $\Tr(\phi_1\phi_1)$ operator at finite coupling. We plot ${\rm Im}\; \Delta$ as a function of the coupling $g$. Purple dots show numerical data obtained by solving the QSC. We also show the analytic weak coupling result to order $g^2$, $g^6$, $g^8$ and $g^{10}$ as the curves colored blue, yellow, green and red correspondingly. The higher the order, the better they agree with data at weak coupling, while the opposite is true for large coupling.}
    \label{fig:plot}
\end{figure}

In table \ref{numrestab} we present a subset of our numerical data, also shown on figure \ref{fig:plot}. We show the results for the state $\Delta_+$ for which ${\rm Im}\;\Delta>0$ (for the other state $\Delta$ simply has opposite imaginary part). We took $\kappa=e^i$ and $\hat\kappa=e^{i/(1+\zeta_3)}$ in order to avoid any accidental relations between the parameters. We also found that the real part of $\Delta$ is equal to $2$ with very high accuracy, so we only give the results for the imaginary part.

\begin{table}[h]
\begin{center}
\begin{tabular}{|c|c|c|c|}
\hline
$g$ & ${\rm Im}\;\Delta$ & $g$ & ${\rm Im}\;\Delta$ \\
\hline
 0.10 & 0.0294933409 & 0.90 & 1.34367362 \\
 0.20 & 0.116344200 & 0.95 & 1.42921633 \\
 0.30 & 0.251836247 & 1.00 & 1.51293336 \\
 0.40 & 0.420163976 & 1.05 & 1.59485349 \\
 0.50 & 0.605259408 & 1.10 & 1.67501243 \\
 0.60 & 0.795473687 & 1.15 & 1.75345580 \\
 0.70 & 0.983949715 & 1.20 & 1.83023548 \\
 0.80 & 1.16719228 & 1.25 & 1.90540742 \\   
%  0.1 & 0.0294933409115 \\
%  0.2 & 0.116344200310 \\
%  0.3 & 0.251836247437 \\
%  0.4 & 0.420163975606 \\
%  0.5 & 0.605259408495 \\
%  0.6 & 0.795473686851 \\
%  0.7 & 0.983949714611 \\
%  0.8 & 1.16719228289 \\
%  0.9 & 1.34367361984 \\
%  1.0 & 1.51293336131 \\
%  1.1 & 1.67501243122 \\
%  1.2 & 1.83023547966 \\
 \hline
\end{tabular}
\end{center}
\caption{Numerical data for the scaling dimension at finite coupling from the Quantum Spectral Curve.}
\label{numrestab}
\end{table}

At weak coupling we can test our analytic 5-loop prediction \eq{weakcoupdelta} against the numerical data. Choosing a particular small value of the coupling $g=0.1$, we computed the anomalous dimension with a high precision and found $\Delta = 2+0.02949334091154i$. In Table~\ref{numantab} we compare it with the analytic perturbative result where we include more and more orders at weak coupling. It is clear that the more terms we add, the better our analytic result matches the numerical value\footnote{Adding the $g^4$ term does not improve the accuracy since it is equal to zero.}. This also serves as a nontrivial consistency check of our perturbative calculation.

\begin{table}[h]
\begin{center}
\begin{tabular}{|c|c|c|c|}
\hline
order & ${\rm Im}\; \Delta^{\rm perturbative}$ & ${\rm Im}\; \Delta^{\rm numerical}$ & $|{\rm difference}|$\\
\hline
$g^2$  & 0.029530412 & 0.029493341 &  0.000037071 \\
$g^4$  & 0.029530412 & 0.029493341 &  0.000037071 \\
$g^6$  & 0.029488834 & 0.029493341 &  0.000004507 \\
$g^8$  & 0.029493865 & 0.029493341 &  0.000000524 \\
$g^{10}$  & 0.029493275 & 0.029493341 &  0.000000065
\\ \hline
\end{tabular}
\end{center}
\caption{Comparison between the 5-loop analytic weak coupling prediction and the numerical data for the anomalous dimension at $g=0.1$ with $\kappa=e^i$ and $\hat\kappa=e^{i/(1+\zeta_3)}$.}
\label{numantab}
\end{table}

At strong coupling it becomes rather time-consuming to get a high-precision result as one should include many terms in the expansion \eq{pasgen}-\eq{fg_ansatz} of the $\bP$-functions which are moreover expanded in powers of $x$ and not $x^2$, increasing by several times the number of needed terms compared to simpler states like Konishi. In a subsequent version of the paper we will extend the range and precision of our numerical data, as well as present the fit at strong coupling. At the moment we cannot say conclusively what is the scaling at strong coupling, with two possibilities being $\Delta\sim g$ (i.e. $\sim \lambda^{1/2}$ like e.g. for the generalized cusp dimension \cite{Drukker:2011za}) or $\Delta\sim g^{1/2}$ (i.e. $\sim \lambda^{1/4}$ like for short operators such as Konishi \cite{Gubser:1998bc}). We should note that the state we are considering is not a conventional string theory state since its energy $\Delta$ is not real. It would be very interesting to see if string theory methods nevertheless can provide a prediction for the strong coupling expansion coefficients.

\section{Conclusions}
\label{sec:concl}

In this paper we have shown that the integrability framework works perfectly for the short $J=2$ operator in $\gamma$-deformed SYM that has caused some controversy in the past. We have computed its scaling dimension to 5 loops analytically and at finite coupling numerically, finding full agreement with known predictions.

It would be important to clarify the string theory description of the state we consider, which seems somewhat unconventional as its dimension is not real. Similar `tachyonic' states have been discussed previously in e.g. \cite{Dymarsky:2005nc,Pomoni:2008de} (see also \cite{Tseytlin:1999ii,Klebanov:1999ch,Klebanov:1999um}), and perhaps one could also establish connections with the proposed dual models to the fishnet theory \cite{Gromov:2019aku,Basso:2018agi,Basso:2019xay}. Like for the Konishi state, a re-expansion of semiclassical results may allow one to get a prediction for the first few strong coupling coefficients.
%A numerical solution of the QSC will clarify the scaling of the energy at strong coupling, and we will report on this in the susbsequent version of the paper once we extend our numerical data.

An interesting future direction is to extend our results to  operators with bare dimension $\Delta=4,6,8,\dots$ that are related by analytic continuation in $g$ to the state we considered.  They all share the same quantum numbers (except $\Delta$) and correspond to excited states in the integrability description. In the fishnet model their spectrum has been computed exactly \cite{Grabner:2017pgm,Kazakov:2018gcy}. The QSC should allow one to directly implement the analytic continuation via complex values of the coupling and study these states analytically as well as numerically. Similar states were explored for the cusped Wilson lines in \cite{Grabner:2020nis,Cavaglia:2018lxi}. One may expect that for some of these states our $c_{a,n}$ coefficients in the QSC will become divergent at weak coupling like in \cite{Grabner:2020nis} (which is however just a technical obstacle). It also remains an interesting problem to reproduce the exact results for the spectrum in the general version of the fishnet theory via the QSC, as well as to use the QSC to do a controlled expansion around the fishnet limit.

Our perturbative results also give new data for the double-trace operator beta functions. It would be interesting to study other short operators in the $\gamma$-deformed model and generate more results of this kind, perhaps establishing new links between diagrammatic computations and integrability. Such links are also aided by the presence of twists whcih give extra parameteres distinguishing various Feynman graphs\footnote{For an early example of predictions for diagrams from integrability in the $\beta$-deformation see \cite{Gromov:2010dy}}. A curious fact we have already found is that the $g^4$ term of the scaling dimension vanishes which calls for an explanation at the level of diagrams.

As we have seen that the QSC works very well for the $\gamma$-deformation, it would be important to use it for other other deformed models such as those with twists in AdS (see \cite{Cavaglia:2020hdb} for recent progress). One should also be able to extend the QSC to the dipole deformation  which should also be integrable \cite{Guica:2017jmq} and occupies an intermediate place between them and the $\gamma$-deformation.

\section*{Acknowledgements}

We thank M. Alfimov, J. Caetano, A. Cavaglia, G. Ferrando, D. Grabner, N. Gromov, Julius, V. Kazakov, A. Tseytlin, E. Vescovi and K. Zarembo for related discussions. We are also grateful to N. Gromov, V. Kazakov and G. Sizov for sharing preliminary parts of the Mathematica code for the numerical solution, and to N. Gromov for access to computational facilities. We thank V. Kazakov for collaboration at early stages of this project. F.L.-M. acknowledges funding by the LabEx ENS-ICFP: ANR-10-LABX-0010/ANR-10-IDEX-0001-02 PSL*. M.P. work is supported by the grant "Exact Results in Gauge and String Theories" from the Knut and Alice Wallenberg foundation.

\appendix

\section{QSC details}
\label{app:qsc}
Here we present some technical details on the Quantum Spectral 
Curve.

In order to ensure correct asymptotics of all the Q-functions, one should impose several nontrivial relations on the $c_{a,n}$ coefficients entering the $\bP$-functions. They are rather lengthy and we give them in a Mathematoca file accompanying this paper. As an example, the relation between $\Delta$ and the coefficients of the $\bP$-functions reads (it was first presented in \cite{Gromov:2017cja}):
\beqa\label{eqDelta}\nonumber
&&(\Delta-2)^2=\!-\!\left[{\left(\kappa
   -\hat{\kappa }\right)^2 \hat{\kappa } \left(\hat{\kappa }+1\right) \left(\kappa
   \hat{\kappa }-1\right)^2}\right]^{-1}\!\left[
        -2 g^8 \left(\kappa \!-\!\hat{\kappa }\right)^2 \left(\hat{\kappa }\!-\!1\right)^2
   \left(\hat{\kappa }\!+\!1\right) \left(\kappa  \hat{\kappa }\!-\!1\right)^2 c_{4,1}^2
\right.+\\ \nonumber
&&\left.
   +2 g^6
   \left(\kappa -\hat{\kappa }\right)^2 \left(\hat{\kappa }-1\right)^2 \left(\hat{\kappa
   }+1\right) \left(\kappa  \hat{\kappa }-1\right)^2 c_{4,2}
\right.+\\ \nonumber
&&\left.
   -2 i g^4 \left(\kappa
   -\hat{\kappa }\right) \left(\hat{\kappa }-1\right) \hat{\kappa } \left(\kappa
   \hat{\kappa }-1\right) \left(\left(\hat{\kappa }^2+1\right) \kappa ^2-4 \hat{\kappa }
   \kappa +\hat{\kappa }^2+1\right) c_{4,1}
\right.+\\ \nonumber
&&\left.
   -2 i \left(\kappa ^2-1\right) \left(\kappa
   -\hat{\kappa }\right) \left(\hat{\kappa }-1\right)^2 \hat{\kappa } \left(\hat{\kappa
   }+1\right) \left(\kappa  \hat{\kappa }-1\right) c_{3,-1}
\right.+\\
&&\left.
   +2 \left(\kappa -\hat{\kappa
   }\right)^2 \left(\hat{\kappa }-1\right)^2 \left(\hat{\kappa }+1\right) \left(\kappa
   \hat{\kappa }-1\right)^2 c_{2,0}
\right.+\\ \nonumber
&&\left.
   -2 \left(\hat{\kappa }+1\right) \left(\hat{\kappa }
   \left(-\left(\hat{\kappa }^3+\hat{\kappa }\right) \kappa ^4-\left(\hat{\kappa
   }-3\right) \left(3 \hat{\kappa }-1\right) \left(\hat{\kappa }^2+1\right) \kappa ^3
\right.\right.\right.+\\ \nonumber
&&\left.\left.\left.
   -2
   \left(\hat{\kappa } \left(\hat{\kappa } \left(\left(\hat{\kappa }-7\right) \hat{\kappa
   }+18\right)-7\right)+1\right) \kappa ^2
\right.\right.\right.+\\ \nonumber
&&\left.\left.\left.
   -\left(\hat{\kappa }-3\right) \left(3
   \hat{\kappa }-1\right) \left(\hat{\kappa }^2+1\right) \kappa -\hat{\kappa }
   \left(\hat{\kappa }^2+1\right)\right)-2 g^2 \left(\kappa -\hat{\kappa }\right)^2
   \left(\hat{\kappa }-1\right)^2 \left(\kappa  \hat{\kappa }-1\right)^2\right)
   \right]
   \eeqa
\noindent   
The coefficients entering the fourth order equation \eq{bax5} on $\bQ_i$ read:
\beq
D_0={\rm det}
\(
\bea{llll}
\bP^{1[+2]}&\bP^{2[+2]}&\bP^{3[+2]}&\bP^{4[+2]}\\
\bP^{1}&\bP^{2}&\bP^{3}&\bP^{4}\\
\bP^{1[-2]}&\bP^{2[-2]}&\bP^{3[-2]}&\bP^{4[-2]}\\
\bP^{1[-4]}&\bP^{2[-4]}&\bP^{3[-4]}&\bP^{4[-4]}
\eea
\)
\eeq

\beq 
D_1={\rm det}
\(
\bea{llll}
\bP^{1[+4]}&\bP^{2[+4]}&\bP^{3[+4]}&\bP^{4[+4]}\\
\bP^{1}&\bP^{2}&\bP^{3}&\bP^{4}\\
\bP^{1[-2]}&\bP^{2[-2]}&\bP^{3[-2]}&\bP^{4[-2]}\\
\bP^{1[-4]}&\bP^{2[-4]}&\bP^{3[-4]}&\bP^{4[-4]}
\eea
\)
\eeq

\beq
D_2={\rm det}
\(
\bea{llll}
\bP^{1[+4]}&\bP^{2[+4]}&\bP^{3[+4]}&\bP^{4[+4]}\\
\bP^{1[+2]}&\bP^{2[+2]}&\bP^{3[+2]}&\bP^{4[+2]}\\
\bP^{1[-2]}&\bP^{2[-2]}&\bP^{3[-2]}&\bP^{4[-2]}\\
\bP^{1[-4]}&\bP^{2[-4]}&\bP^{3[-4]}&\bP^{4[-4]}
\eea
\)
\eeq

\beq
\bar D_1={\rm det}
\(
\bea{llll}
\bP^{1[-4]}&\bP^{2[-4]}&\bP^{3[-4]}&\bP^{4[-4]}\\
\bP^{1}&\bP^{2}&\bP^{3}&\bP^{4}\\
\bP^{1[+2]}&\bP^{2[+2]}&\bP^{3[+2]}&\bP^{4[+2]}\\
\bP^{1[+4]}&\bP^{2[+4]}&\bP^{3[+4]}&\bP^{4[+4]}
\eea
\)
\eeq

\beq
\bar D_0={\rm det}
\(
\bea{llll}
\bP^{1[-2]}&\bP^{2[-2]}&\bP^{3[-2]}&\bP^{4[-2]}\\
\bP^{1}&\bP^{2}&\bP^{3}&\bP^{4}\\
\bP^{1[+2]}&\bP^{2[+2]}&\bP^{3[+2]}&\bP^{4[+2]}\\
\bP^{1[+4]}&\bP^{2[+4]}&\bP^{3[+4]}&\bP^{4[+4]}
\eea
\)
\eeq

\bibliographystyle{nb}

\bibliography{biblio_gammadef}

%\begin{thebibliography}{99}
 
%\bibitem{Marboe:2019wyc}
 % C.~Marboe and E.~Widén,
  %``The fate of the Konishi multiplet in the $\beta$-deformed Quantum Spectral Curve,''
%  JHEP {\bf 2001} (2020) 026
%  doi:10.1007/JHEP01(2020)026
%  [arXiv:1902.01248 [hep-th]].

%\end{thebibliography}
\end{document}